\documentclass[a4,12pt]{article}
\begin{document}
\setlength{\parindent}{15pt}
\setlength{\baselineskip}{18pt}
\setlength{\parskip}{10pt}
\def\thefootnote{\fnsymbol{footnote}}

\noindent  
{\bf \large Proper Time for Spin $1/2$ Particles}

\vskip 5mm

\noindent
Shoju Kudaka $^1$ and Shuichi Matsumoto $^2$\footnote{E-mail: shuichi@edu.u-ryukyu.ac.jp }\\
{\it $^1$Department of Physics, University of the Ryukyus, Okinawa 903-0213, 
Japan}\\
{\it $^2$Department of Mathematics, University of the Ryukyus, Okinawa 903-0213, Japan}

\vskip 1cm
We find a quantum mechanical formulation of proper time for spin $1/2$ particles within the framework of the Dirac theory. It is shown that the rate of proper time can be represented by an operator called the \lq \lq tempo operator'', and that the proper time itself be given by the integral of the expectation value of the operator. The tempo operator has some terms involving the Pauli spin matrices, and the evolution of the proper time is influenced by the spin state via these terms. The relation between the tempo operator and the metric tensor is elucidated.

\vskip 10mm
\section{Introduction}

Proper time is one of the most important concepts when we consider a particle in a gravitational field. In Einstein's theory of general relativity, the total of all events is taken to make up a Riemannian manifold with a metric tensor $g_{\mu \nu }$; the proper time of a particle is defined as the length of its orbit, and is interpreted as the time read by a clock attached to the particle. The so-called time delay, which is predicted from this interpretation, is now established through such phenomena as particle collapse and radiation redshift. 
  
  In this geometrical description, however, a particle is treated as a single geometrical point: It should be noted that the proper time of a particle is defined only in terms of its coordinates $x^{\mu }$;  
\begin{equation}
d\tau _{\rm cl}={\sqrt {g_{\mu \nu }dx^{\mu }dx^{\nu }}}. \label{eq:classicaleq}
\end{equation}
On the other hand, a {\it physical particle} may have other degrees of freedom, such as spin, in addition to its position. Therefore the question is (Q-1) whether or not the rate of proper time is affected by the spin and if it is, then (Q-2) how should the classical formula (\ref{eq:classicaleq}) be modified to account for it. In this article we try to answer these questions, focusing our attention on a spin $1/2$ Dirac particle.  

  The problem of a small classical spinning object in a gravitational field has been tackled by many authors, and the equations of motion have been found in various forms\ \cite{Math, Papa, Dix}. It is, moreover, known that there exists a coupling of quantum spin to space-time curvature which causes a deviation from geodesic trajectories\ \cite{Anan}. It is natural, then, to consider that the velocity of a particle is affected by such a coupling as well. Therefore, judging from the classical formula (\ref{eq:classicaleq}), we are led to the possibility that the rate of proper time is affected by the spin; this is the background to the questions (Q-1) and (Q-2).

  In this article, in order to reach a quantum theoretical formulation of proper time for spin $1/2$ particles, we consider two different routes: One starts with the Dirac equation in curved space-time, rewriting it as a Schr{\"o}dinger equation, and then elucidating the influence of the spin on the velocity of the particle. Another approach starts with a paticular invariant integral, one which is a natural starting point when we deal with the proper time from the point of view of working with invariance under all general coordinate transformations and all local Lorentz ones. These two routes finally converge on an identical definition for the proper time of the particle. This seems to the authors to indicate that there is some truth to the formulation thus obtained.

\vskip 5mm
\section{Spin $1/2$ particle in a gravitational field} 

  We consider a spin $1/2$ particle of mass $m$ in curved space-time with a metric tensor $g_{\mu \nu }$. The Dirac equation for spinor $\psi $ is given by
\begin{equation}
i\gamma ^jv_j^{\mu }D_{\mu }\psi -m\psi =0, \label{eq:kisoEq}\end{equation}
which is covariant for an arbitrary local Lorentz transformation. In Eq. (\ref{eq:kisoEq}), $\gamma ^j$ are the Dirac $\gamma $ matrices\ \cite{B-D}, and $v_j^{\mu }$ a vierbein\ \cite{MTW} by which the metric tensor is related to $\eta _{i j}$: 
$$v_i^{\mu }v_j^{\nu }g_{\mu \nu }=\eta _{i j} \ \ (={\rm diag.}(1, -1, -1, -1)).$$
$D_{\mu }$ denotes covariant derivatives 
$$D_{\mu }\equiv \partial _{\mu }+{1\over 2}\omega _{i j, \mu }S^{i j}, $$ 
where the spin connection is given by   
\begin{equation}
\omega _{i j, \mu }=v_i^{\nu }\left( \partial _{\mu }v_{j \nu }-\{ ^{\lambda }_{\nu \mu }\} v_{j \lambda }\right) ,\label{eq:4kasetu}
\end{equation}
and $S^{i j}$ is the generator of the Lorentz group; $S^{i j}\phi =0$ for a scalar, $S^{i j}\psi =(1/4)[\gamma ^i, \gamma ^j]\psi $ for a spinor, and $(S^{i j}A)^k=\eta ^{i k}A^j-\eta ^{j k}A^i$ for a vector. Using equations (\ref{eq:kisoEq}) and (\ref{eq:4kasetu}), we can show the continuity equation
\begin{equation}
\partial _{\mu }\left( {\sqrt {-g}}\ {\overline \psi }\gamma ^jv_j^{\mu }\psi \right) =0  \hskip 1cm ({\overline \psi }\equiv \psi ^{\dagger }\gamma ^0). \label{eq:conservation}
\end{equation}
 
  In this article, the gravitational field is assumed to be static in the sense that we can select a coordinate frame $(x^{\mu })$ in which all components $g_{\mu \nu }$ do not depend on $t\equiv x^0$. Moreover we restrict ourselves to weak-field conditions, so that 
$$g_{\mu \nu }=\eta _{\mu \nu }+h_{\mu \nu }\hskip 1cm (h_{\mu \nu }\ll 1). $$

  If we expand Einstein's field equation for vacuum gravitational field in powers of $h_{\mu \nu }$, and if we keep only linear terms, then we get equations 
\begin{equation}
\Delta h_{\mu \nu }=0, \label{eq:EinsteinEq}
\end{equation}
where we have assumed the De Donder's coordinate condition $\partial _{\mu }\{ {\sqrt {-g}}g^{\mu \nu }\} =0$; this condition means, under our assumptions, that 
\begin{equation}
\sum _{j=1}^3\partial _jh_{0 j}=0, \hskip 1cm \sum _{j=1}^3\partial _jh_{i j}+{1\over 2}\partial _ih=0\hskip 3mm (i=1, 2, 3),
\label{eq:zahyojyoken}
\end{equation}
where $h\equiv \eta ^{\mu \nu }h_{\mu \nu }$. In the following calculations, we use a vierbein
\begin{equation}
v_i^{\mu }=\pmatrix{ 1-h_{0 0}/2 & h_{0 1} & h_{0 2} & h_{0 3} \cr 0 & 1+h_{1 1}/2 & h_{1 2}/2 & h_{1 3}/2 \cr 
0 & h_{2 1}/2 & 1+h_{2 2}/2 & h_{2 3}/2 \cr
0 & h_{3 1}/2 & h_{3 2}/2 & 1+h_{3 3}/2 \cr  
} \label{eq:vierbein}
\end{equation}
with the subscript $i$ denoting the row index and the superscript $\mu $ the column index.

\vskip 5mm
\section{Schr{\"o}dinger equation}

  Using equations (\ref{eq:4kasetu})、(\ref{eq:EinsteinEq}), (\ref{eq:zahyojyoken}) and (\ref{eq:vierbein}), we can rewrite equation (\ref{eq:kisoEq}) in the form  
\begin{equation}
i\partial _t\psi =H\psi , \label{eq:Schro}
\end{equation}
where 
%\begin{equation}
$$H=m\beta +{\cal O}+{\cal E}$$
%\end{equation}
%\label{eq:Hdef}
and
$${\cal O}=(1+\phi ){\bf \alpha }\cdot {\bf p}, \hskip 5mm {\cal E}=m\beta \phi -{1\over 4}(\nabla \times {\bf g})\cdot {\bf \sigma }-{\bf g}\cdot {\bf p};$$
we set 
$$\phi \equiv h_{0 0}/2, \hskip 1cm {\bf g}\equiv (-h_{0 1}, -h_{0 2}, -h_{0 3}),$$ 
and
\begin{equation}
p_j\equiv -iv_j^{\mu }\partial _{\mu }-{i\over 8}(\partial _jh)\hskip 5mm(j=1, 2, 3). \label{eq:pjdef}
\end{equation}
In the following calculations, we use an explicit representation in which the matrices are 
$$\alpha _j=\pmatrix{0 & \sigma _j \cr \sigma _j & 0 \cr } \hskip 1cm \beta =\pmatrix{ 1 & 0 \cr 0 & -1 \cr }$$
where the $\sigma _j$ are the familiar $2\times 2$ Pauli matrices and the unit entries in $\beta $ stand for $2\times 2$ unit matrices. 

  The operators $p_j$ defined by (\ref{eq:pjdef}) are self adjoint with respect to the inner product
%\begin{equation}
$$\langle \psi _1\vert \psi _2\rangle _0\equiv \int d^3{\bf x}{\sqrt {-g}}\psi _1^{\dagger }\psi _2,$$
%\label{eq:4naiseki} \end{equation}
and satisfy the commutation relations
$$[p_j, p_k]={1\over 2}\sum _{l=1}^3(-\partial _jh_{k l}+\partial _kh_{j l})\partial _l.$$
The operator $H$ is self adjoint with respect to the inner product 
\begin{equation}
\langle \psi _1\vert \psi _2\rangle \equiv \int d^3{\bf x}{\sqrt {-\ ^3g}}\psi _1^{\dagger }\psi _2; \label{eq:3naiseki}
\end{equation}
$^3g$ is defined by  
$$^3g\equiv \det \pmatrix{ g_{1 1} & g_{1 2} & g_{1 3} \cr g_{2 1} & g_{2 2} & g_{2 3} \cr g_{3 1} & g_{3 2} & g_{3 3} \cr }=-1+h_{1 1}+h_{2 2}+h_{3 3},$$
and we have 
\begin{equation}
{\sqrt {-g}}=(1+\phi ){\sqrt {-\ ^3g}}. \label{eq:g-3g}
\end{equation}
Taking (\ref{eq:vierbein}) and (\ref{eq:g-3g}) into account, we have  
$${\sqrt {-g}}\ {\overline \psi }\gamma ^jv_j^0\psi ={\sqrt {-g}}(1-\phi ){\overline \psi }\gamma ^0\psi ={\sqrt {-\ ^3g}}\psi ^{\dagger }\psi ,$$
so we can rewrite equation (\ref{eq:conservation}) in the form
%\begin{equation}
$$\partial _t\left( {\sqrt {-\ ^3g}}\psi ^{\dagger }\psi \right) =-\sum _{\mu =1}^3\partial _{\mu }\left( {\sqrt {-g}}\ {\overline \psi }\gamma ^jv_j^{\mu }\psi \right) .$$
%\label{eq:rewrited} \end{equation}
In this article, we content ourselves with a non-relativistic particle and argue about the quantum mechanics governed by the Schr{\"o}dinger equation (\ref{eq:Schro}); the probability density is given by 
$${\sqrt {-\ ^3g}}\psi ^{\dagger }\psi .$$

  The presence of the odd term ${\cal O}$, coupling the large and small components of the Dirac spinor, necessitates the Foldy-Wouthuysen (FW) transformation\ \cite{FW}. After four FW transformations we find 
\begin{eqnarray*}
UHU^{\dagger }&=&m\beta +m\beta \phi -{1\over 4}(\nabla \times {\bf g})\cdot \sigma -{\bf g}\cdot {\bf p}+{1\over {2m}}\beta (1+\phi ){\bf p}^2\\
&&-{1\over {4m}}\beta (\nabla \phi \times \sigma )\cdot {\bf p}+{1\over {4m}}\beta \sum _{i, j, k, l=1}^3\epsilon _{i j k}(\partial _ih_{j l})p_l \sigma _k\\
&&+{1\over {16m^2}}\sum _{i, j, k, l=1}^3\epsilon _{i j k}\{ (\partial _jg_l+\partial _lg_j)p_lp_i+p_lp_i(\partial _jg_l+\partial _lg_j)\} \sigma _k
\end{eqnarray*}
to the order of $1/m^2$ where $U$ denotes the product of those four FW transformations.

  For a given $\psi $ we define $\psi _{FW}$ and two-component wave functions $\Psi $ and $\chi $ by
$$\psi _{FW}\equiv U\psi =\pmatrix{ \Psi \cr \chi \cr }.$$
We now assume that $\psi $ is a superposition of positive eigenstates of $H$. Then we have 
\begin{equation}
\vert \chi /\Psi \vert =O(1/m^3) \label{eq:hikaku}
\end{equation}
and therefore, for a non-relativistic particle, the equation (\ref{eq:Schro}) reduces to 
\begin{equation}
i\partial _t\Psi =H_{FW}\Psi ;  \label{eq:reduced}
\end{equation}
the Hamiltonian $H_{FW}$ is given by  
\begin{eqnarray*}
H_{FW}&\equiv &m+m\phi -{1\over 4}(\nabla \times {\bf g})\cdot \sigma -{\bf g}\cdot {\bf p}+{1\over {2m}}(1+\phi ){\bf p}^2\\
&&-{1\over {4m}}(\nabla \phi \times \sigma )\cdot {\bf p}+{1\over {4m}}\sum _{i, j, k, l=1}^3\epsilon _{i j k}(\partial _ih_{j l})p_l \sigma _k\\
&&+{1\over {16m^2}}\sum _{i, j, k, l=1}^3\epsilon _{i j k}\{ (\partial _jg_l+\partial _lg_j)p_lp_i+p_lp_i(\partial _jg_l+\partial _lg_j)\} \sigma _k
\end{eqnarray*}
to the order of $1/m^2$.

  Using Hamiltonian $H_{FW}$, we can show that  
\begin{eqnarray}
{\dot x}^i&=&i[H_{FW}, x^i] \nonumber \\
&=&-g_i+{1\over m}(1+\phi )p_i+{1\over {4m}}\sum _{j=1}^3(p_jh_{i j}+h_{i j}p_j) \nonumber \\
&&-{1\over {4m}}(\nabla \phi \times {\bf \sigma })_i-{1\over {4m}}\sum _{j, k, l=1}^3\epsilon _{j k l}(\partial _jh_{i l})\sigma _k \nonumber \\
&&+{1\over {16m^2}}\sum _{j, k, l=1}^3\epsilon _{i j k}\{ (\partial _jg_l+\partial _lg_j)p_l+p_l(\partial _jg_l+\partial _lg_j)\} \sigma _k \nonumber \\
&&+{1\over {16m^2}}\sum _{j, k, l=1}^3\epsilon _{j k l}\{ (\partial _jg_i+\partial _ig_j)p_l+p_l(\partial _jg_i+\partial _ig_j)\} \sigma _k  \label{eq:sokudo} 
\end{eqnarray}
for $i=1, 2, 3$. Pauli matrices $\sigma _k$ are involved in the right-hand side: This indicates that the velocity of a particle is indeed affected by its spin state as presumed in the introductory section.  
\vskip 5mm
\section{An invariant integral}

  Now we turn our attention to the integral 
\begin{equation}
\int _{\Omega }d^4x{\sqrt {-g}}\psi (x)^{\dagger }\beta \psi (x), \label{eq:fuhenint}
\end{equation}
where $\Omega $ denotes an arbitrary domain in space-time. This integral is invariant under not only any arbitrary general coordinate transformation but also any arbitrary local Lorentz transformation; the uniqueness of this integral will be discussed in the concluding section from the point of view of working with those invariances. 

  Let $\Omega $ be of the form 
$$\Omega =[t_0, t]\times {\bf R}^3$$
and define 
%%\begin{equation}
$$\tau (t)\equiv \int _{\Omega }d^4x{\sqrt {-g}}\psi (x)^{\dagger }\beta \psi (x)=\int _{t_0}^tdt\int d^3{\bf x}{\sqrt {-g}}\psi (x)^{\dagger }\beta \psi (x), $$
%%\label{eq:tautdef} \end{equation}
then we have 
%%\begin{equation}
$$\tau (t)=\int _{t_0}^tdt\langle \psi \vert (1+\phi )\beta \vert \psi \rangle $$
%%\nonumber \label{eq:QproperT} \end{equation}
by using the inner product (\ref{eq:3naiseki}).

  The rate $d\tau /dt$ can be expressed by
\begin{equation}
{d\over {dt}}\tau (t)=\langle \psi \vert (1+\phi )\beta \vert \psi \rangle =\langle \psi _{FW}\vert U(1+\phi )\beta U^{\dagger }\vert \psi _{FW}\rangle \label{eq:tempoorig} 
\end{equation}
and we find, to the order of $1/m^2$,  
\begin{eqnarray}
U(1+\phi )\beta U^{\dagger }&=&(1+\phi )\beta -{1\over {2m^2}}\beta (1+\phi ){\bf p}^2 \nonumber \\
&&+{1\over {4m^2}}\beta (\nabla \phi \times {\bf \sigma })\cdot {\bf p}+{1\over {4m^2}}\beta \sum _{i, j, k, l=1}^3\epsilon _{j k l}(\partial _jh_{i l})\sigma _kp_i\nonumber \\
&&+\ {\rm odd\ terms\ with\ the\ order}\ \geq 1/m. \label{eq:UbetaU}
\end{eqnarray}
Substituting (\ref{eq:UbetaU}) into (\ref{eq:tempoorig}), we have 
\begin{eqnarray}
&&{d\over {dt}}\tau (t)=\langle \Psi \vert 1+\phi -{1\over {2m^2}}
(1+\phi ){\bf p}^2 \nonumber \\
&&\hskip 1cm +{1\over {4m^2}}(\nabla \phi \times {\bf \sigma })\cdot 
{\bf p}
+{1\over {4m^2}}\sum _{i, j, k, l=1}^3\epsilon _{j k l}
(\partial _jh_{i l})\sigma _kp_i\vert \Psi \rangle 
\label{eq:tempohyoji} 
\end{eqnarray}
to the order of $1/m^2$, where \lq \lq odd terms'' in (\ref{eq:UbetaU}) can be neglected by using (\ref{eq:hikaku}).
The operator occurring in the right-hand side of the 
above equation (\ref{eq:tempohyoji}) will be denoted by ${\cal T}$ and will be called the \lq \lq tempo operator'': 
\begin{eqnarray}
{\cal T}&\equiv &1+\phi -{1\over {2m^2}}(1+\phi ){\bf p}^2  \nonumber \\
&&+{1\over {4m^2}}(\nabla \phi \times {\bf \sigma })\cdot {\bf p}+{1\over {4m^2}}\sum _{i, j, k, l=1}^3\epsilon _{j k l}(\partial _jh_{i l})\sigma _kp_i. \label{eq:tempoop}
\end{eqnarray}
Using (\ref{eq:EinsteinEq}), we can easily see that 
\begin{eqnarray*}
{\cal T}^2&=&1+2\phi -{1\over {m^2}}(1+2\phi ){\bf p}^2+{i\over {m^2}}\sum _{j=1}^3(\partial _j\phi )p_j \nonumber \\
&&+{1\over {2m^2}}(\nabla \phi \times {\bf \sigma })\cdot {\bf p}+{1\over {2m^2}}\sum _{i, j, k, l=1}^3\epsilon _{j k l}(\partial _jh_{i l})\sigma _kp_i.  \nonumber %\label{eq:T2}
\end{eqnarray*} 

  Now, we proceed to the final step of our calculations: We pay attention to the equation  
$${{d\tau _{cl}}\over {dt}}={\sqrt {g_{\mu \nu }{{dx^{\mu }}\over {dt}}{{dx^{\nu }}\over {dt}}}},$$
one which is obtained from the classical formula (\ref{eq:classicaleq}). We replace the $c$-numbers $dx^i/dt \ (i=1, 2, 3)$ in the equality
$${{dx^{\mu }}\over {dt}}g_{\mu \nu}{{dx^{\nu }}\over {dt}} =1+2\phi -\sum _{j=1}^3\left( g_j{{dx^j}\over {dt}}
+{{dx^j}\over {dt}}g_j\right) -\left({{d{\bf x}}\over {dt}}\right) ^2
+\sum _{i, j=1}^3{{dx^i}\over {dt}}h_{i j}{{dx^j}\over {dt}}$$
with the operators ${\dot x}^i$ given by (\ref{eq:sokudo}). Then, using (\ref{eq:EinsteinEq}) and (\ref{eq:zahyojyoken}), we can show that  
$${\dot x}^{\mu }g_{\mu \nu }{\dot x}^{\nu }={\cal T}^2,$$
where we set ${\dot x}^0=1$. Thus we have proved 
$${\cal T}={\sqrt {{\dot x}^{\mu }g_{\mu \nu }{\dot x}^{\nu }}},$$
which is the goal of our calculations. 

\vskip 5mm
\section{Concluding remarks}

  Our conclusions may be summarized as follows:
\begin{enumerate}
\item A quantity $\tau (t)$ with parameter $t$ can be defined from an integral (\ref{eq:fuhenint}) which is invariant with respect to all general coordinate and local Lorentz transformations. 
\item The rate $d\tau /dt$ is represented as an expectation value of an operator ${\cal T}$ called the tempo operator.  
\item On the other hand, the Dirac equation in curved space-time reduces to the equation (\ref{eq:reduced}) from which we can get the velocity operator ${\dot {\bf x}}$. 
\item Finally, we can show that 
\begin{equation}
{\cal T}={\sqrt {{\dot x}^{\mu }g_{\mu \nu }{\dot x}^{\nu }}}
.  \label{eq:Tsqrt}
\end{equation}
\end{enumerate}
The authors consider that the above conclusion strongly suggests that this quantity $\tau (t)$ should be interpreted as the proper time for the Dirac particle; the classical formula (\ref{eq:classicaleq}) should be limited to the case of a scalar particle.  

  The tempo operator ${\cal T}$ given by (\ref{eq:tempoop}) has some terms involving the Pauli spin matrices $\sigma _k$. The evolution of the proper time is influenced by the spin state via these terms; this is our answer to (Q-1) in the introduction. The equation (\ref{eq:Tsqrt}) means moreover that this influence is of such a form that the classical formula (\ref{eq:classicaleq}) can still survive if we reinterpret the formula as an equation between operators. The metric tensor seems to maintain its important role in quantum theory: This is our answer to the question (Q-2).
  
  In the following, we add some remarks. First, we have to estimate the strength of the interaction between the rate of proper time and the spin state. A typical term to be estimated is     
$${{\hbar }\over {4m^2c^2}}(\nabla \phi \times {\bf \sigma })\cdot {\bf p}.$$
The magnitude of this term becomes non-negligible only when the gravitational field varies notably over the Compton wave length $h/mc$ of the particle and when its velocity approaches the speed of light. Therefore, as far as laboratory experiments are concerned, the new effect is so small that it is very difficult to be measured. In this sense, our conclusion is open to the criticism that it is not verifiable. However, the authors think that this conclusion necessarily follows from accepting the Dirac equation (\ref{eq:kisoEq}). We hope that the importance of this is recognized and that, for example, the implications of the tempo operator are duly considered.  

  Secondly, we have to explain why we may regard the rate of proper time to be expressed by an operator：
\begin{enumerate}
\item In the theory of relativity, the proper time of a particle depends on its history, in the sense that it is determined by the orbit in space-time. On the other hand, the rate of proper time does not depend on the history.
\item In the quantum mechanics of a single particle, an operator can describe a physical quantity at $t=$constant but cannot, however, represent any quantity which depends on the history of the particle. 
\item Therefore, if we could define \lq \lq the quantum mechanical proper time'' of a particle in the Dirac theory, it would be natural to think that the rate of proper time can be represented by an operator, and that the proper time itself be given by the integral of the expectation value of the operator. 
\end{enumerate}

  Some investigations start from the postulate that proper time and rest mass are operators which satisfy a commutation relation\ \cite{Gr, Kal, K-M}. These approaches lead to some desirable results, but at the same time face a fundamental difficulty: If we let an operator correspond to the coordinate time $t$ such that it satifies the relation $[t, H]=i$ with the Hamiltonian $H$, then the spectrum of the Hamiltonian has to be continuous; this was proved by Pauli\ \cite{Pauli}. This manifestly contradicts the existence of a discrete energy spectrum. Similarly, if we assume that the proper time $\tau $ and rest mass m are operators which satisfy the relation $[\tau , m]=i$, we are led to a result that conflicts with the existence of a discrete mass spectrum. In this article, we have succeeded to find a quantum mechanical formulation of proper time within the Dirac theory for relativistic quantum mechanics. We did not require any additional assumptions such that the rest mass be represented by an operator. It has been shown that an operator corresponds to the rate $d\tau /dt$ and that the proper time itself can be represented through the expectation value of this operator. Our method does not involve any logical difficulty.    

  Next, we should explain the reason why we selected the integral (\ref{eq:fuhenint}) as our starting point. We assume for a while space-time to be flat and begin by finding a matrix $D$ for which the integral 
$$\int _{t_0}^tdt\langle \psi \vert D\vert \psi \rangle =\int _{t_0}^tdt\int d^3{\bf x}\psi (x)^{\dagger }D\psi (x)$$
can be interpreted as \lq \lq the quantum mechanical proper time of the particle'' which passes in the course of an interval $[t_0, t]$ of coordinate time. In order to refer to \lq \lq proper time'', this integral has to be invariant under all Lorentz transformations
$$x'^j={L^j}_kx^k.$$
But to require that this invariance be represented by the equation
\begin{equation}
\int _{t_0}^tdt\int d^3{\bf x}\psi (x)^{\dagger }D\psi (x)=
\int _{t'_0}^{t'}dt'\int d^3{\bf x}'\psi '(x')^{\dagger }D\psi '(x') \label{eq:tooformal}
\end{equation}
is too formal. This is because Eq.(\ref{eq:tooformal}) has no meaning, as the interval $[t_0', t']$ cannot be determined from the interval $[t_0, t]$. Accordingly, the authors decided to represent this invariance by the equation 
\begin{equation}
\int _{\Omega }d^4x\psi (x)^{\dagger }D\psi (x)=\int _{\Omega '}d^4x'\psi '(x')^{\dagger }D\psi '(x'), \label{eq:kakikae}
\end{equation}
where $\Omega $ denotes an arbitrary domain in space-time and $\Omega '$ its image under the Lorentz transformation. 

  If we take into account not only proper Lorentz transformations but also space reflection, then we can deduce from (\ref{eq:kakikae}) that the matrix $D$ must be $\beta $ multiplied by a real constant. That is to say, our formulation for proper time necessarily leads us to the intgral 
$$\int _{\Omega }d^4x\psi (x)^{\dagger }\beta \psi (x).$$
If we bring this integral into the curved space-time in the simplest way, we get the integral (\ref{eq:fuhenint}). It is invariant under not only local Lorentz transformations but also general coordinate ones. 

  Finally, we refer to other literature in which the Dirac matrix $\beta $ plays an essential role: From the pioneer work of Fock\ \cite{Fock}, much effort has been made to introduce explicitly the concept of an invariant parameter into the relativistic quantum mechanics\ \cite{Aparicio, Gynga}. In the majority of such contributions, the proper time is a $c$-number parameter and various proper time derivatives have been proposed. For example, Ellis\ \cite{Ellis} defines, following Corben\ \cite{Cor}, a proper time derivative of operator by
$${{dX}\over {d\tau}}=\beta {{dX}\over {dt}},$$
and shows that this method enables us to derive some well known results for the Dirac equation in a comparatively effortless manner. There the Dirac matrix $\beta $ is taken to connect proper time and coordinate time rates of change of operators. 

  There exists a clear difference between such approaches and ours presented in this article: Our attempt has been founded on the view that the rate $d\tau /dt$ is a measurable quantity and should be associated with an operator. The authors hope that this approach can cultivate a better understanding of time.

\vfill
\end{document}